\documentstyle[12pt]{article}
\begin{document}
\baselineskip 1.7\baselineskip
\centerline
{\bf Analytic Treatment of Mott-Hubbard Transition in the Half-Filled Hubbard
Model}
\vspace{1cm}
\centerline{Jongbae Hong and Hae-Young Kee}
\centerline{Department of Physics Education, Seoul National University,
Seoul 151-742, Korea}

\vspace{2cm}

\centerline{\bf ABSTRACT}
\vspace{1cm}

 The Mott-Hubbard transition in the half-filled Hubbard model is studied
analytically for the paramagnetic ground state and the classical
N\'{e}el state.  The single-particle density of states is obtained by
calculating the Green's function represented by the infinite continued
fraction. The paramagnetic solution shows that the Mott-Hubbard transition is
signaled by both collapsing Hubbard bands and appearing $\delta$-function
peak at midgap, and the transition is second order at the ground state. We
also provide specific heats in metallic regime to demonstrate that our results
recover those of infinite dimension.

\vspace{2cm}


\eject

 The Hubbard model has been studied first by Hubbard[1], Gutzwiller[2], and
 Kanamori[3] as a model for a narrow band system. It has been recently
 revisited in relation to a possible model for high-Tc superconductors[4,5].
 Besides cuprate superconductors, the Hubbard model may also be applicable
 to describing the metal-insulator transition(MIT) in the materials like
 $V_{2-y}O_3$[6] in which paramagnetic insulator, paramagnetic metal,
 antiferromagnetic insulator, and a new phase antiferromagnetic metal[7] are
 clearly separated in the temperature-pressure and temperature-filling
 factor($y$) phase diagram. Transitions between these phases[8], mass
 enhancement[9] due to strong correlation and magnetic ordering[9,10] are
 common phenomena of the Mott-Hubbard systems such as $Sr_{1-x}La_xTiO_3$[11],
 $Ni(Se_{1-x}S_x)_2$[12] as well as $V_2O_3$.  In addition, an interesting
 search for a superconducting transition in the metallic $V_2O_3$ is suggested
 in close relation to a Mott-Hubbard insulator shown in the rare-earth oxide
 superconductors and the strong electron correlation shown in the
 heavy-fermion superconductors.[7] These bring renewed interest in the
 Mott-Hubbard transition(MHT), recently.

   The MHT accompanies both electronic and magnetic transitions. Pure $V_2O_3$
 in a paramagnetic metal phase becomes antiferromagnetic insulator through the
 antiferromagnetic ordering of the vanadium spins at a temperature around
 150K.[6] Above this temperature, paramagnetic metal-insulator transition
 occurs. This transition temperature, however, can be suppressed until very
 low temperature by applying hydrostatic pressure, adjusting the oxygen
 stoichiometry, and alloying with titanium sesquioxide.[7-9]  Therefore, both
 paramagnetic metal-insulator transition due to strong correlation and
 paramagnetic metal-antiferromagnetic insulator transition due to
 antiferromagnetic ordering are interesting.  These transitions accompany
 Fermi liquid quasiparticle mass enhancement through the strong electron
 correlations, which has been studied by many experiments[8,9,11]. On the
 other hand, no  successful theories exist to explain the experimental results
 except the large-d approach[12-15] as far as we know.

  The beauty of large-d approach lies in the ability to solve the Hubbard
model, in principle, exactly in a well-defined limit in which spatial
correlations do not play an important role.  Unfortunately, some degree of
insight into the local dynamics contributing  to the end-results is concealed
as a component of the numerical  self-consistency that is an  integral part of
this approach. In an effort to compliment this procedure, in this Letter we
introduce an extremely simple approach to the dynamics of the MHT by
calculating the single-particle  density of states(DOS) through a continued
fraction formalism for the Green's function. The formalism is known as
projection operator technique[16], recurrence relation method[17], and
dynamical Lanzcos method[5].  The method that we use has been previously
applied with great success to the response function of the electron gas at a
metallic density at which usual perturbation technique may not be that
successful.[18] From the study  on  the electron  gas,  we  found that
infinite  continued  fraction provides analytic  expression  for the
dynamics  of  the electron  system  by  using minimal amount of information
about the ground-state correlations.

    Here we study the MHT on a bipartite lattice for the paramagnetic ground
 state and the classical N\'{e}el state. The paramagnetic metal-insulator
transition is realized when temperature is  above  the N\'{e}el temperature.
Below the  N\'{e}el temperature, however,  the MHT does  not occur at a
finite on-site Coulomb interaction $U$  on a bipartite lattice because of
 appearing  antiferromagnetic long-range  order due to perfect nesting.[19]
Even though the paramagnetic  metal-insulator transition is realized  only
above the N\'{e}el temperature, the physics of the transition in the
artificial paramagnetic ground  state, for  instance, the Brinkman-Rice
scenario[20], is also interesting to understand the strong correlation
effect in the MHT in the Hubbard  model. This  problem, especially  in a
finite dimension,  has been considered as one of quite difficult tasks in
condensed matter physics, because one must treat both spatial correlations
and local fluctuations equally rigorously.

   As a lot of insight is gained by reproducing complicated results within a
very simple theory, in this work we show that a very large fraction of the
essential physics described in the large-d approach such as the Hubbard bands,
the quasi-particle peak, and the second order feature of the MHT at ground
state can be reproduced by a simple continued fraction expansion about atomic
state. This reproduction may be understood by the arguments advanced as part
of the large-d approach that a great deal of physics in the MHT is dominated
by local dynamics.[12-14]

    As for a finite dimension, however, no first principles calculation has
been done to obtain the single-particle DOS even for the atomic state as far
as we know. But our analytic solution keeps the dimensional dependence even
though approximation becomes cruder as one goes to a lower dimension. The
effect of spatial correlation which brings more prominent dimensional
dependence to the DOS will be studied in the future.

      In this Letter, we first show that the classical N\'{e}el state
which is close to the real ground state of the half-filled Hubbard model
yields Hubbard gap for any $U>0$ in the single-particle DOS. Thus we recover a
well-known fact that the MHT occurs at $U=0$ for the Hubbard model on a
bipartite lattice. The paramagnetic solution for the DOS shows the Hubbard
feature[1] and the Brinkman-Rice feature[20] for the MHT simultaneously in a
very simple analytic form. We also show that the thermodynamics obtained by
this DOS reproduces that of large-d, which confirms the similarity of the
approximations used in the large-d approach and our simple method.

   The single-particle DOS is obtained by calculating the one-particle Green's
function of the fermion operator at the same site, i.e.,
$\langle\Psi_0|\{c^{\dagger}_{j\sigma}(t),c_{j\sigma}\}|\Psi_0\rangle$ where
$c_{j\sigma}^{\dagger}$ and $c_{j\sigma}$ are fermion creation and
annihilation operators with spin $\sigma$ at site $j$, curly brackets means
anticommutator, and $|\Psi_0\rangle$ denotes ground state. The DOS
$\rho_{\sigma}(\omega)$ is given by[1]
\begin{equation}
\rho_{\sigma}(\omega)=\frac{2}{N}\lim_{\epsilon\rightarrow 0^+}\sum_j
{\rm Im} G_{jj}^{(-)}(\omega-i\epsilon),
\end{equation}
where
\begin{eqnarray}
G_{jj}^{(-)}(\omega-i\epsilon)&=     &\frac{i}{2\pi}
\int_0^{\infty}\langle\Psi_0|\{c_{j\sigma}^{\dagger}(t),c_{j\sigma}\}|\Psi_0
\rangle e^{-i\omega t-\epsilon t}dt \nonumber \\
                              &=     &\frac{1}{2\pi}\langle\Psi_0|
\{c_{j\sigma},
(\omega-L-i\epsilon)^{-1}c_{j\sigma}^{\dagger}\}|\Psi_0\rangle \nonumber \\
                              &\equiv&\frac{i}{2\pi}a_0(z)|_{z=i\omega+
\epsilon},
\end{eqnarray}
where $L$ is the Liouville operator. The superscript $(-)$ denotes usual
notation of the advanced Green's function.  The Green's function (2) may be
represented by an infinite continued fraction[5,16,17]. Thus the
single-particle DOS is written as
\begin{equation}
\rho_{\sigma}(\omega)=\frac{1}{\pi N}\lim_{\epsilon\rightarrow
0^+}\sum_j{\rm Re}a_0(z)|_{z=i\omega+\epsilon},
\end{equation}
where
\begin{equation}
a_0(z)=\frac{1}{z-\alpha_0+\frac{\Delta_1}{z-
\alpha_1+\frac{\Delta_2}{z-\alpha_2+\ddots}}},
\end{equation}
where
$\alpha_{\nu}=(iLf_{\nu},f_{\nu})/(f_{\nu},f_{\nu})$,
$\Delta_{\nu}=(f_{\nu},f_{\nu})/(f_{\nu-1},f_{\nu-1})$. The inner product is
defined by $(A,B)=\langle\Psi_0|\{A,B^{\dagger}\}|\Psi_0\rangle$ where $A$ and
$B$ are operators of the Liouville space, $B^{\dagger}$ is the adjoint of $B$.

  The Hubbard model which we study here is written as
\begin{equation}
H=-\sum_{<jl>\sigma}t_{jl}c_{j\sigma}^{\dagger}c_{l\sigma}+\frac{U}{2}\sum_
{j\sigma}n_{j\sigma}n_{j,-\sigma},
\end{equation}
where $<jl>$ means nearest  neighbor sites. To obtain the DOS
$\rho_{\sigma}(\omega)$ where $\sigma$ denotes spin-up state, we choose
$f_0=c_{j\sigma}^{\dagger}$ and calculate $\alpha_{\nu}$ and $\Delta_{\nu}$
by using the recurrence relation[17]
\begin{equation}
f_{\nu+1}=iLf_{\nu}-\alpha_{\nu}f_{\nu}+\Delta_{\nu}f_{\nu-1}
\end{equation}

  The key approximation in this process is the Hartree-Fock approximation in
calculating the ground state correlation function such as $\langle n_{j,-
\sigma} n_{l,-\sigma}\rangle =\langle n_{j,-\sigma}\rangle\langle n_{l,-
\sigma} \rangle$. This corresponds to setting the ground state
$|\Psi_0\rangle$ as a product of atomic states and to neglecting spatial
correlations which is valid at large-d limit. A similar decoupling scheme has
been done on the average containing the operator $[H,n_{j\sigma}]$ whose
average vanishes, since no steady current exists in the ground state.

  First we obtain $\rho_{\sigma}(\omega)$ for the classical N\'{e}el state to
study the MHT for a state with antiferromagnetic long-range order.  We take
antiferromagnetic spin configuration such that the spin at site $j$ is
$\downarrow$. Then we obtain
\begin{eqnarray}
\alpha_{\nu} &=& iU \; \; \mbox{when $\nu$ is odd}, \nonumber \\
             &=& 0  \; \; \; \mbox{when $\nu$ is even},
\end{eqnarray}
and
\begin{equation}
\Delta_1=qt^2, \; \; \Delta_{\nu}=(q-1)t^2 \; \; \mbox{for $\nu\geq 2$}.
\end{equation}
To obtain $\rho_{-\sigma}(\omega)$, on the other hand, we choose
$f_0=c_{j,-\sigma}^{\dagger}$. Then we obtain $\alpha_{\nu}=0$ for odd $\nu$
and $iU$ for even $\nu$ and have the same $\Delta_{\nu}$ as above.  Using
these, $\rho(\omega)=\rho_{\sigma}(\omega)+\rho_{-\sigma}(\omega)$ is given by
\begin{equation}
\rho(\omega)=\frac{\frac{4}{\pi}q(q-1)t^4|\omega|\sqrt{\left(\omega^2-
\frac{U^2}{4}\right)
\left(\frac{U^2+16(q-1)t^2}{4}-\omega^2\right)}}{(q-2)^2\left(\omega^2-
\frac{U^2}{4}\right)^2+
q^2t^4\left(\omega^2-\frac{U^2}{4}\right)\left(\frac{U^2+16(q-1)t^2}{4}-
\omega^2\right)}
\end{equation}
Eq. (9) satisfies sum rule exactly and yields Hubbard gap which makes the
system insulator for any $U>0$.  We plot Eq. (9) for $U=0$ and $U=1t$ for
$q=4$ in Fig. 1, for instance.

     For the paramagnetic ground state in which one can see the MHT at finite
$U$, we obtain the following regular patterns for $\Delta_{\nu}$ and
$\alpha_{\nu}$ for large-$U$:
\begin{eqnarray*}
\Delta_1       &=&\frac{U^2}{4}+qt^2, \\
\Delta_2       &=&\frac{\frac{1}{2}U^2qt^2+q(q-1)t^4}{\frac{1}{4}U^2+qt^2}, \\
\Delta_{2\nu+1}&=&\frac{U^2}{4}\{1+O(\frac{t^2}{U^2})\}\equiv a, \\
\Delta_{2\nu+2}&=&4qt^2\{1+O(\frac{t^2}{U^2})\}\equiv b,
\end{eqnarray*}
where $\nu\geq 1$ and $\alpha_0=\alpha_1=\alpha_2=iU/2$ and
$\alpha_{\nu}=i\frac{U}{2}\{1+O(t^2/U^2)\}$ for $\nu\geq 3$. Only even powers
of $t/U$ appear in the expansion for $\Delta_{\nu}$ and $\alpha_{\nu}$ for
$\nu\geq 3$. A further approximation $(q-1)\approx q$ which is valid at higher
dimension has also been made.

  By neglecting terms of $O(t^2/U^2)$, the infinite continued fraction (4) can
be easily calculated, and the result is
\begin{equation}
a_0(\tilde{z})=\frac{\tilde{z}+\frac{\Delta_2}{2b}\left[\frac{(b-a)}
{\tilde{z}}-\tilde{z}+\frac{1}{\tilde{z}}\sqrt{(\tilde{z}^2+a-b)^2+4b
\tilde{z}^2}\right]}{\tilde{z}^2+\frac{\Delta_2}{2b}\left[(b-a)-\tilde{z}^2
+\sqrt{(\tilde{z}^2+a-b)^2+4b\tilde{z}^2}\right]+\Delta_1},
\end{equation}
where $\tilde{z}=z-\alpha_0$.  We choose  positive sign  in front of  the
square  root instead of  negative sign to  satisfy  the boundary  condition
$a_0(t=0)=1$,  where $a_0(t)$  is the inverse Laplace transform of
$a_0(\tilde{z})$.

   If  one  sets the chemical potential at $\mu=\frac{U}{2}$, Eq. (3)
gives the DOS as follows:
\begin{eqnarray}
\rho_{\sigma}(\omega)&=&\frac{\frac{\Delta_1\Delta_2}{2b\pi|\omega|}
                        \sqrt{W}}{\left[\frac{\Delta_2}{2b}(b-a)+\Delta_1
                        +(\frac{\Delta_2}{2b}-1)\omega^2\right]^2+\left[
                        \frac{\Delta_2^2}{4b^2}W\right]} \;\;\; \mbox{for
                        a$>$b} \\
                     &=&\frac{\left(1-\frac{a}{b}\right)\delta(\omega)}
                        {\frac{\Delta_1}{\Delta_2}+ \left(1-
                        \frac{a}{b}\right)}+ \nonumber \\
                     & &\frac{\frac{\Delta_1\Delta_2}{2b\pi|\omega|}\sqrt{W}}
{\left[\frac{\Delta_2}{2b}(b-a)+\Delta_1+(\frac{\Delta_2}{2b}-1)\omega^2
\right]^2+\left[\frac{\Delta_2^2}{4b^2}W\right]} \;\;\; \mbox{for a$<$b},
\end{eqnarray}
where $W=\{\omega^2-(\sqrt{a}-\sqrt{b})^2\}\{(\sqrt{a}+\sqrt{b})^2-
\omega^2\}$. From the single-particle DOS, one can see that MIT occurs at
$a=b$, i.e., $U_c=4\sqrt{q}t$ and the Fermi liquid quasi-particle mode appears
at midgap in metallic phase, i.e., $U<U_c$ and the Hubbard bands have band
width $4\sqrt{q}t$ in insulator($a>b$) and $U$ in metallic phase($a<b$). The
band gap is $|U-4\sqrt{q}t|$. To draw $\rho_{\sigma}(\omega)$ in the large-d
limit, we scale $t$ as $t=t_{\ast}/\sqrt{2q}$ and set $t_{\ast}=1$.  We plot
three typical cases, i.e., (a) $U=4$(two bands), (b) $U=U_c=2\sqrt{2}$ (single
band), (c) $U=2.4$(two bands with $\delta$-function peak at $\omega=0$) in
Fig. 2.

   The $\delta$-function quasi-particle peak at zero frequency is of course an
artifact, and a more realistic calculation would lead to a dispersive band. In
fact, a similar calculation to present one but expanded for small-$U$ shows
the dispersive band, which will be submitted in a separated paper.
Nevertheless, the weight of the $\delta$-function peak gives us a measure of
the quasi-particle mass renormalization $Z_F=m/m^{\ast}$. The quasi-particle
mass diverges as $U$ approaches  $U_c$  from  metallic  side. The
Brinkman-Rice  scenario  for MIT[20] can be seen here.

   It is interesting to compare our results to those of large-d. Our DOS
shows that the MHT in paramagnetic ground state is second order, which
coincides with the recent results[15] of infinite dimension. The dispersive
band shown in the large-d results may be reproduced by applying the Luttinger
theorem[21] which fixes the height of the quasi-particle peak to the value of
the DOS at Fermi energy of the noninteracting system. For further comparison
with the large-d approach, we extract a part of thermodynamics from the DOS
(12).

    The extension of our formalism to the finite temperature is rather simple,
since the temperature-dependence only enters in the correlation function such
as $\langle n_{j-\sigma}n_{l-\sigma}\rangle$ which vanishes within
Hartree-Fock approximation. Therefore, the temperature-dependence does not
appear in the DOS in atomic limit. Thus the internal energy has temperature-
dependence only in the Fermi distribution function $f(\omega)$. In the
metallic regime, the internal energy is given by
\begin{equation}
E(T)=\int_{-\infty}^{\infty}\omega f(\omega)\{\rho_{inc}(\omega)+
\rho_{qp}(\omega)\}d\omega ,
\end{equation}
where $\rho_{inc}(\omega)$ is the incoherent part of Eq. (12) and
$\rho_{qp}(\omega)$ is the quasi-particle DOS. The latter is the coherent part
of Eq. (12) divided by the quasi-particle mass renormalization $Z_F$, which
can be written as
\begin{equation}
\rho_{qp}(\omega)=\frac{1}{\pi Z_F}\frac{\eta}{\omega^2+\eta^2}.
\end{equation}
The parameter $\eta$ is determined by the condition $\rho_{qp}(0)=N(0)/Z_F$,
where $N(0)=\sqrt{2}/\pi$[22]. The specific heats calculated from Eq. (13) for
$U$'s which are a little bit less than $U_c$ are shown in Fig. 3. One can
see a peak at low-temperature regime which is given by the Fermi liquid
quasi-particle.

   One of interesting features which looks peculiar among the results of
large-d approach is shown in the specific heat at metallic regime($U<U_c$).
The specific heats for different $U$'s show a crossing at a certain
temperature, $T\approx 6.5$.  This phenomenon has also been shown in the
experiments for \ $^3{\rm He}$[23].  But the physical origin of it is not
clear yet.  These phenomena are the characteristics of large-d result[8].
Other thermodynamic quantities obtained from Eqs. (11) and (12) will be
appeared in the forthcoming paper.

 As final remarks, we mention about our approximation. The Hartree-Fock
approximation in calculating the static correlation functions is the only
approximation when we obtain the DOS for the classical N\'{e}el state. For the
paramagnetic solution, however, we neglect $O(t^2/U^2)$ terms compared with
unity in obtaining $\Delta_{\nu}$ and $\alpha_{\nu}$ for $\nu\geq 3$ in
addition to the Hartree-Fock approximation. The neglected terms may shifts the
positions of poles a little bit when they are included. We imagine that this
effect may not change the over-all physics contained in the DOS. Even under
these approximation, our DOS satisfies sum rule exactly, i.e.  $\int_{-
\infty}^{\infty}\rho_{\sigma} (\omega)d\omega=1$, for $U$'s given in Fig. 2.

  The underlying reason for the remarkable agreement between the large-d and
the simple continued fraction method must lie in the fact that the physics of
the large-d limit is dominated by the atomic limit. The continued fraction
expansion used here may provide the possibility to treat the effects of
quantum fluctuation in the classical N\'{e}el state and spatial correlation as
well as doping. These will be the subjects of forthcoming paper.

\vspace{1cm}

 The authors  wish to  express their gratitude  to the International  Center
for Theoretical Physics for  financial support and hospitality. They  also
thank Prof. P. Coleman, Prof. P. Fazekas for useful discussions and Prof.  Yu
Lu for encouragement. One of authors(J.H.) also thanks the Government of
Japan for support.  This work has  been supported by SNU-CTP and Basic Science
Research Institute Program, Ministry of Education.

\eject

\noindent [1] J. Hubbard, Proc. Roy. Soc. {\bf A276}, 238 (1963) and
{\bf A281}, 401 (1964).

\noindent [2] M. Gutzwiller, Phys. Rev. Lett. {\bf 10}, 159 (1963).

\noindent [3] J. Kanamori, Prog. Theor. Phys. {\bf 30}, 275 (1963).

\noindent [4] P. W. Anderson, Science {\bf 235}, 1196 (1987).

\noindent [5] E. Dagotto, Rev. Mod. Phys. {\bf 66}, 763 (1994).

\noindent [6] D. B. McWhan, A. Menth, J. P. Remeika, W. F. Brinkman, and T.
M. Rice, Phys. Rev. B {\bf 7}, 1920 (1973).

\noindent [7]   S.  A. Carter,  T.   F. Rosenbaum,  J.   M. Honig,  and   J.
Spalek, Phys. Rev. Lett. {\bf 67}, 3440 (1991).

\noindent [8] S. A. Carter {\it et  al}., Phys.  Rev. B  {\bf 43}, 607
(1991).

\noindent [9] S. A. Carter {\it et  al}., Phys.  Rev. B  {\bf 48}, 16841
(1993).

\noindent [10] W. Bao {\it  et al}., Phys. Rev. Lett. {\bf 71}, 766 (1993).

\noindent [11] Y. Tokura Phys. Rev. Lett. {\bf 70}, 2126 (1993).

\noindent [12] A.   Georges and   G. Kotliar,   Phys.  Rev.  B   {\bf 45},
6479 (1992).

\noindent [13] A. Georges and W. Krauth, Phys. Rev. B {\bf 48}, 7167
(1993).

\noindent [14] M. J. Rozenberg, G. Kotliar, and X. Y. Zhang, Phys. Rev.
B {\bf 49}, 10181 (1994).

\noindent [15] M. Caffarel and W. Krauth, Phys. Rev. Lett. {\bf 72},
1545 (1994); M. J. Rozenberg, G. Moeller, and G. Kotliar,
SISSA \#cond-mat/9402056

\noindent [16] H. Mori, Prog. Theor. Phys. {\bf 33}, 423 (1965), and {\bf 34},
399 (1965); P. Fulde, {\it Electron Correlations in Molecules and Solids}
Solid-State Sciences Vol. 100 (Springer-Verlag, Berlin, 1993).

\noindent [17] J.  Hong,   J. Kor.  Phys.  Soc. {\bf  20},  174  (1987); M.
H.    Lee,   J. Math.   Phys. {\bf  24},    2512 (1983),   Phys. Rev.  Lett.
{\bf 49}, 1072 (1982), and Phys. Rev.  B{\bf 26} (1982).

\noindent [18] J.  Hong and  M. H.  Lee, Phys.  Rev. Lett.  {\bf 70},  1972
(1993); {\bf 55}, 2375 (1985).

\noindent [19] J. C. Slater,  Phys. Rev.   {\bf  82}, 538 (1951);  J.   E.
Hirsch and S. Tang, Phys. Rev. Lett. {\bf 62}, 591 (1989).

\noindent [20] W. F. Brinkman and T. M. Rice, Phys. Rev. B {\bf 2}, 4302
(1970).

\noindent [21] E. Muller-Hartmann, Z. Phys. B {\bf 74}, 507 (1989).

\noindent [22] The continued fraction (4) for small-$U$ limit gives
$\rho_{\sigma}(0)=\sqrt{q-1}/\pi qt$ which corresponds to $\sqrt{2}/\pi
t_{\ast}$.

\noindent [23] D. S. Greywall, Phys. Rev. B {\bf 27}, 2747 (1983).

\eject

{\bf Figure Captions}

\noindent Fig. 1 : The single-particle density of states $\rho(\omega)$ for
the classical N\'{e}el state.  The dotted line is for $U=0$ and the solid line
for $U=1t$.

\noindent Fig. 2 : The single-particle density of states
$\rho_{\sigma}(\omega)$ of the paramagnetic phase for large-d limit. (a)
$U=4t_{\ast}$, (b) $U=U_c=2\sqrt{2}t_{\ast}$, (c) $U=2.4t_{\ast}$. The number
(0.11) in (c) represents the  weight of $\delta$-function.

\noindent Fig. 3 : The specific heats for $U<U_c=2\sqrt{2}$. The solid,
dashed, and dotted lines are for $U$=2.6, 2.4, and 2.2, respectively. We set
$t_{\ast}=1$.

\end{document}